 \documentclass[aps,pre,twocolumn,groupedaddress]{revtex4}
\usepackage[dvips]{graphicx}
\usepackage{times}

\begin{document}

%\preprint{}

\title{ Conductance of single-atom magnetic junctions: A first-principles study }


\author{ Yi-qun Xie$^{1,2}$ }
\email[]{ yqxie@shnu.edu.cn}
\author{ Qiang Li$^{1}$, Lei Huang$^{1}$, Xiang Ye$^{1}$ }
%\thanks{}
\affiliation{$^{1}$Department of Physics, Shanghai Normal University, Shanghai 200232,
People's Republic of China.}
\affiliation{$^{2}$Beijing Computational Science Research Center, 3 Heqing Road,
Beijing 100084, People's Republic of China.  }

\author{San-Huang Ke$^{3}$}
\email[]{ shke@tonji.edu.cn}
\affiliation{ $^{3}$Key Laboratory of Advanced Microstructured Materials, MOE,
Department of Physics, Tongji University, 1239 Siping Road, Shanghai 200092,
People's Republic of China.}
%\affiliation{$^{2}$Beijing Computational Science Research Center, Beijin 100084, People¡¯s Republic of China. }


\date{\today}

\begin{abstract}
We present a first-principles investigation to show that the contact conductance of a
half conductance quantum ($G_0/2$) found previously does not generally hold for single-atom
magnetic junctions composed of a tip and an adatom adsorbed on a surface. The
contact conductance of the Ni-Co/Co(111) junction is approximately
$G_0/2$, while for the Co-Co/Co(111), Ni-Ni/Ni(111), and Ni-Ni/Ni(001) junctions
the contact conductances  are  0.80$G_0$, 1.55$G_0$, and 1.77$G_0$, respectively. The
deviation from $G_0/2$ is mainly caused by the variation of the spin-down
conductance largely determined by the minority $d$ orbitals, as the spin-up one
changes little for different junctions.

\end{abstract}

%\keywords{}

\maketitle

%\section{Introduction}

The electrical transport property of the atomic contact has attracted much
attention for its enormous potential in future atomicscale electronic devices.
The conductance of a spin-degenerated atomic-size contact is quantized in units of
$G_0=2e^2 /h$,  where $e$ is the proton charge and $h$ is
the Planck's constant according to the Landauer's formula\cite{Landauer}.
In the case of magnetic systems the spin degeneracy is
removed, and each spin-polarized channel can contribute
up to $G_0/2$ to the total conductance.

The conductance of atomic junctions has been extensively studied by experimental
techniques such as the scanning tunneling microscope (STM)
\cite{stm1,stm2,stm3,stm4,stm5}, mechanically controlled break junctions (MBCJ) technique [7-11], and
electrochemical methods \cite{ElChem2,ElChemNi}. Conductances of fractional and
integer $G_0$ have been reported for both nonmagnetic and magnetic junctions.
Of the particular interest is the observation of $G_0/2$ conductance, e.g., for
the noble Pt, Pd, and magnetic Co single-atom chains \cite{Ugarte2},  as well as Cu and Ni
nanowires \cite{ElChemNi,CuHalfQuantum}, although the controversy still exists
by arguing that the contamination of $H_2$ may cause the $G_0/2$
conductance \cite{Untiedt}.
Theoretically, the transport property of atomic junctions has been investigated
intensively by using either tight-binding methods \cite{TBAu,TB2,TB3Ni} or
density functional theory (DFT) combined with quantum transport calculations based on
nonequilibrium Green's function (NEGF) formalism [18-27]. However, theoretical
calculations have hardly obtained the $G_0/2$ conductance so far. This may be
attributed to the difference between the simulation model and the real junction
structure which is usually difficult to know experimentally.

Recently, a STM  experiment found that in the contact regime the conductance is
$G_0/2$ for a single-atom magnetic junction composed of a ferromagnetic Ni
tip in contact with a single Co adatom on a ferromagnetic Co island \cite{stm2}.
This experiment provides a more detailed atomic geometry of the junction than
those ever reported, which enables a precise theoretical modeling to explore the
origin of the $G_0/2$ conductance. Calculations show that the conductance of
$G_0/2$ is not due to a fully polarized single channel but a combination of the
partially open majority channels and the suppressed minority
channels\cite{Tao2}. However, it still remains unknown whether the $G_0/2$
conductance exists generally for other single-atom magnetic junctions, and how
factors like junction structure and species affect the conclusion. In this
work, we perform first-principles calculations to investigate this issue and find that the $G_0/2$
conductance is not a general behavior for several other single-atom magnetic
junctions. This is because the minority-spin conductance is sensitive to the junction
species.

%\section{Computation}

\begin{figure}
\center
\includegraphics[width=8cm]{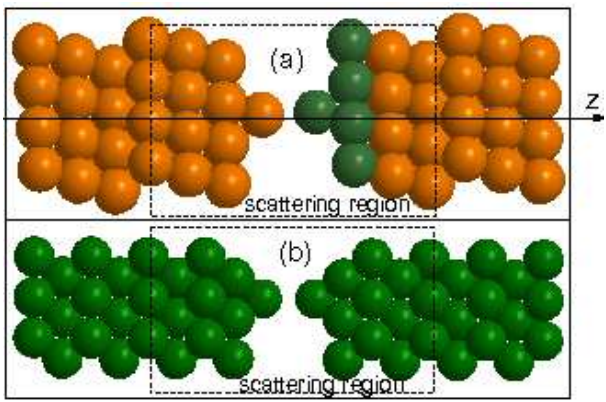}
\caption{ Model system for the tansport calculation of the single-atom magnetic
junction. (a) The Ni-Co/Co(111) junction: A Ni tip (green balls) is placed above the Co adatom on the Co(111)
surface. (b) The Ni-Ni/Ni(001) junction. The transport is along the $Z$ direction. }
\label{fig1}
\end{figure}

We first study the magnetic
junction fabricated by N\'{e}el {\it et al.}\cite{stm2}  The system is modeled
by a tip-adatom/surface junction (denoted by Ni-Co/Co(111)) as displayed in
Fig.\ref{fig1}(a). The single
Co adatom is adsorbed on the $fcc$ site of the $fcc$ Co(111) surface represented by
a three-layer slab, with each layer containing $4\times4$ atoms. The Ni tip is
modeled by a single Ni apex atom and a Ni(111) monolayer on a two-layer Co(111)
slab (test calculations using three Ni(111) layers on a two-layer Co(111) slab
show only minor effect on the result). The
tip-apex atom is placed above the Co adatom in the $z$ direction.  In transport
calculations, these atoms construct the scattering region, and three additional
Co(111) layers are added at the two ends of the scattering region, respectively,
to mimic the left and right electrodes (leads). For the other three junctions considered
in this work, i.e.,
Co-Co/Co(111), Ni-Ni/Ni(111), and Ni-Ni/Ni(001), the treatment is similar except
that a four-layer slab is used for the Ni(001) surface and four additional
atomic layers are added to represent the leads of the Ni(001) junction, as
illustrated in Fig.\ref{fig1}(b). 

In this work, the ferromagnetic spin configuration is
adopted for the whole junction. Our calculations show that if the junction
(including the tip and the substrate) is considered as one system, the ferromagnetic configuration is the ground
state, as was also found in a previous calculation for the Ni-Co/Co(111)
junction \cite{DftAu}. If the junction is considered as two separated
system (i.e., the tip and the substrate) the antiferromagnetic configuration
between them may be also possible, as considered in Ref.\cite{stm3,stm4}. However, the only
consideration of the ferromagnetic configuration does not affect the purpose and
conclusion of this work: The contact conductance of $G_0/2$ is not a general
behavior for different single-atom magnetic junctions.

The scattering region is optimized using VASP code\cite{vasp}. The bottom layer
of the surface and the top layer of the tip are fixed during the structure
optimization while the other atoms are fully
relaxed until the maximum force is smaller than 0.01eV/\AA. Projector augmented-wave
method \cite{paw} is used for the wave function expansion with an
energy cutoff of 300eV. The PW91 version \cite{PW91} of the generalized gradient
approximation (GGA) is adopted for the electron exchange and correlation.

For the quantum transport calculation, we adopt the NEGF-DFT approach \cite{ke2004,datta95} which combines the
NEGF formula for transport with {\it ab initio} DFT calculation for electronic
structure. In practice, the infinitely large open system is divided into three
parts: left lead, right lead, and scattering region, as mentioned above.
The self-consistent Kohn-Sham Hamiltonian of the scattering region and the self-energies
of two semi-infinite leads are used to construct a single-particle Green's
function from which the transmission coefficient ($T$) at any energy is calculated.
The conductance $G$ then follows from a Landauer-type relation.
For the DFT part, we
use a numerical basis set to expand the wavefunction \cite{siesta}: A
double zeta plus polarization basis set (DZP) is adopted for all atomic species.
The Perdew-Burke-Ernzerhof (PBE) \cite{pbe} version of GGA is used for the
electron exchange and correlation and the optimized Troullier-Martins
pseudopotentials \cite{TM} are used for the atomic cores.
We define the spin-polarization ratio at Fermi
energy as $P=(T_\uparrow-T_\downarrow)/(T_\uparrow+T_\downarrow)$, where
$T_\uparrow$ and $T_\downarrow$ denote the transmission coefficient of the
majority and minority spin, respectively.

%\subsection{Results and Discussion}

\begin{figure}
\begin{center}
\includegraphics[width=8cm]{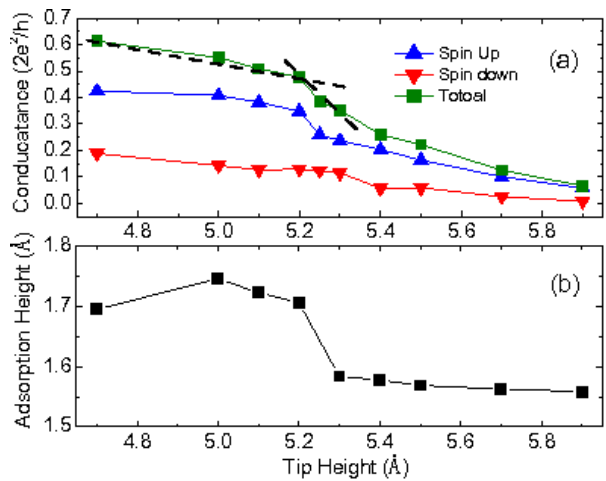}
\end{center}
\caption{(a) The spin-dependent conductance of the Ni-Co/Co(111) junction, and
(b) the adsorption height of the Co adatom relative to the surface for
different tip heights.}
\label{fig2}
\end{figure}

The conductance of the magnetic Ni-Co/Co(111) junction is given in
Fig.\ref{fig2}(a) as a function of the tip height (i.e., the distance between the
tip-apex atom and the surface before the relaxation). It shows that as the
tip height decreases, the conductance increases and shows a faster change in the
transition region around 5.4-5.2\AA, and then increases slowly in the contact
region (below 5.2\AA). The conductance data in the transition and contact regions
can be approximated by two straight lines, and their intersection point defines
the contact conductance, as used in Refs.\cite{stm2,neel3}. According to this
definition, we obtain a contact conductance of 0.48$G_0$ for the Ni-Co/Co(111)
junction, which agrees very well with the experimental result \cite{stm2}.
The two spin components of the conductance give
a spin-polarization ratio of 0.46 for the tip height of 5.2\AA,
which is also in good agreement with the previous first-principles
calculation\cite{Tao2}.

Fig.\ref{fig2}(b) gives the adatom's adsorption height
for different tip heights. On can see that
in the contact region the adsorption height is obviously larger than that in
the tunneling region, and so is the conductance. This is a result of the
enhanced adatom-tip interaction in the contact region.
An important thing to note is that the adsorption height experiences a
large jump for the tip height around 5.3\AA, where the conductance also rapidly increases as
shown in Fig.\ref{fig2}(a).
This indicates that the adsorption height of the adatom has
a direct influence on the conductance because the resulting adatom-tip
separation determines the coupling strength (the degree of orbital overlap). 
Physically, the adsorption height
and the resulting adatom-tip separation for a specific tip height is determined
by the interaction between the tip and the substrate, which may depend on
the junction structure and species. Therefore, to investigate the influence of
these two factors on the conductance is important to find out whether a $G_0/2$
conductance is a general result for other magnetic single-atom junctions.

Next, we investigate the conductance of the following three junctions,
Co-Co/Co(111), Ni-Ni/Ni(111), and Ni-Ni/Ni(001). The first two
have a similar structure as Ni-Co/Co(111) but are of
different species. Thus by comparing their results, we can evaluate the role of
species in determining the conductance of the magnetic junctions. The
Ni-Ni/Ni(001) and  Ni-Ni/Ni(111) junctions are of the same species but have
different structures, which enables us to explore the influence of the atomic
structure on the conductance.

\begin{figure}
\begin{center}
\includegraphics[width=8cm]{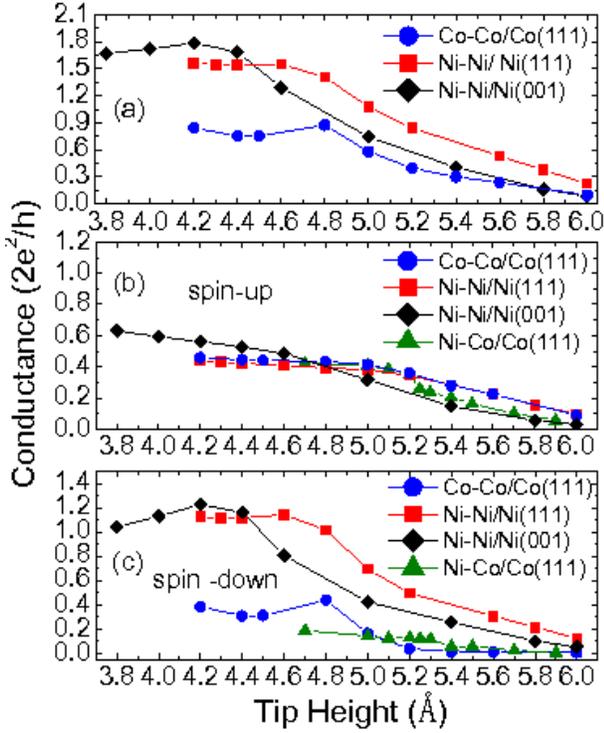}
\end{center}
\caption{(a) Conductance of the Co-Co/Co(111), Ni-Ni/Ni(111), and Ni-Ni/Ni(001)
junctions, respectively.  (b) and (c) show the spin-up and spin-down conductances
of these three junctions and of the Ni-Co/Co(111) junction, respectively.}
\label{fig3}
\end{figure}

Fig.\ref{fig3}(a) displays the conductance as a function of the tip height, and
Figs.\ref{fig3}(b) and (c) give the two spin components of the conductance, for the
Co-Co/Co(111), Ni-Ni/Ni(111) and Ni-Ni/Ni(001) junctions, respectively. We find
that their contact conductances are 0.80$G_0$, 1.55$G_0$, and
1.77$G_0$ with the corresponding spin-polarization ratios of 0.1, -0.45, and -0.33,
respectively. Obviously, the contact conductances of the three junctions deviate
largely from $G_0/2$ and also differ much from each other.

Specifically, the spin-up conductance of the Ni-Co/Co(111), Co-Co/Co(111),
and Ni-Ni/Ni(111) junctions are very close (see Fig.\ref{fig3}(b)), while their
spin-down conductances differ from each other significantly (see
Fig.\ref{fig3}(c)). Since these three junctions are of the similar structure,
this difference is due to the different junction species which give rise to
different local electronic states and tip-adatom couplings.

\begin{figure}
\begin{center}
\includegraphics[width=8cm]{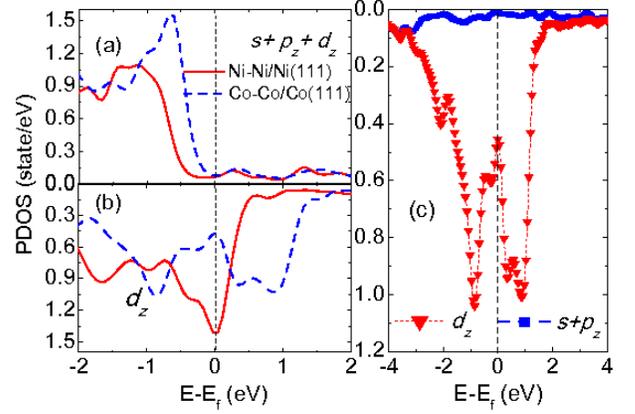}
\end{center}
\caption{(a) The majority and (b) minority PDOS of the Co
adatom in the Co-Co/Co(111) junction and that of the Ni adatom in the
Ni-Ni/Ni(111) junction with a tip height of 4.4\AA. (c) The minority $s$, $p_z$
and $d_z$ PDOS of the Co adatom in the Co-Co/Co(111) junction with 4.4\AA tip
height.}
\label{fig4}
\end{figure}

We find that the spin-up conductance is mainly contributed
by the majority $s$, $p_z$, and $d_z$ (sum of $d_{z^2-r^2}$,
$d_{xz}$ and $d_{yz}$ with $z$ being the transport direction) orbitals and their
contributions to the majority conductance are almost the same for the four
different magnetic junctions \cite{Tao2}. Consequently, the spin-up conductance is
little changed for the different junctions. As an example, we give in Fig.\ref{fig4}(a) the
majority-orbital-projected density of states (PDOS) of the Ni adatom in the
Ni-Ni/Ni(111) junction and that of the Co adatom in the Co-Co/Co(111) junction
with a tip height of 4.4\AA. It can be seen that the majority PDOS
at the Fermi energy are very close for the two adatoms, leading to the similar
spin-up conductances for the two junctions (see Fig.\ref{fig3}(b)).
In contrast, the spin-down conductance is basically determined by the
minority $d_z$ orbitals. Their contribution to the minority PDOS of the adatom at the Fermi
energy is much larger than the $s$ and $p_z$ orbitals, as shown in
Fig.\ref{fig4}(c) for the Co-Co/Co(111) junction. Importantly, as displayed in
Fig.\ref{fig4}(b), the minority $d_z$ PDOS of the adatom at the Fermi energy
varies significantly between the junctions, leading to the very different
spin-down conductances for these two single-atom magnetic junctions.

Now let us have a look at the effect of the junction structure.
By comparing the results for the Ni-Ni/Ni(111) and Ni-Ni/Ni(001) junctions (see
Fig.\ref{fig3}), we find that both the
spin-down and spin-up conductances are different, causing a 0.2$G_0$ difference
in the total conductance as shown in Fig.\ref{fig3}(a).
However, this difference is smaller than those between the Co-Co/Co(111),
Ni-Co/Co(111), and Ni-Ni/Ni(111) junctions caused by the different junction
species (see Fig.\ref{fig3}).
Our results show that the $G_0$/2 conductance does not generally hold for
the single-atom magnetic junctions investigated in this work. This is due to the
the large variation of the spin-down conductance which largely depends on
the junction species as well as the junction structure.

Finally, we would like to comment on the major contribution from the majority or
the minority to the total conductance. In Ref.\cite{stm3}, the authors investigated
the conductance of single-atom magnetic junctions of Cr-Co/Fe(110) and
Cr-Cr/Fe(110). It was found that the conductance is mainly determined
by the majority channel. In this work, we investigate the contact conductance of
single-atom magnetic junctions with different chemical species and structures.
Our calculations show that for the Ni-Co/Co(111)
junction the contact conductance is also mainly determined by
the majority channel: The majority and minority conductances are 0.38 and 0.13
$G_0$, respectively. However, for the other three junctions, the minority
conductance is larger than the majority one. This indicates that the major
contribution to the total conductance, from the majority or the minority, is
also dependent on the junction species and structure.

%\subsection{Conclusion}

In conclusion, by performing first-principles quantum transport calculations we have investigated the
spin-dependent conductance of the four single-atom magnetic junctions. For the
Ni-Co/Co(111) junction the conductance is very close to $G_0/2$, while for the
Co-Co/Co(111), Ni-Ni/Ni(111) and Ni-Ni/Ni(001) junctions the conductances are
far away from $G_0/2$ and differ significantly from each other. It was found
that the spin-up conductance is little influenced by the junction species,
and therefore, this difference is mainly due to the variation of the spin-down component determined by the
minority $d_z$ orbitals which are sensitive to the junction species.
On the other hand, the junction structure can affect both the spin-up and spin-down conductances
but not as remarkably as does the junction species for the spin-down one.
Our calculation shows that the previously found $G_0/2$ conductance does not generally exist for single-atom magnetic
junctions since the conductance is largely dependent on the junction species
and is also influenced considerably by the junction structure.

%\section*{Acknowledgments}
This work was supported by NSFC (Grant No. 51101102 and No. 11174220), Leading
Academic Discipline Project of Shanghai Normal University (Grant
No. DZL712),  and Innovation Researching Fund of Shanghai Municipal
Education Commission (Grant No. 10YZ75 and No. shsf020), as well as by Shanghai
Pujiang Program under Grant No. 10PJ1410000 and the MOST 973 Project under 
Grant No. 2011CB922204.


\end{document}